\documentclass[pra,a4paper,superscriptaddress,twocolumn,showpacs,amsmath,amssymb,floatfix]{revtex4-1}

\usepackage{graphicx}
\usepackage{amssymb}
\usepackage{float}
\usepackage{hyperref}
\hypersetup{
	bookmarks=true,         
	unicode=false,          
	pdftoolbar=true,        
	pdfmenubar=true,        
	pdffitwindow=false,     
	pdfstartview={FitH},    
	pdftitle={My title},    
	pdfauthor={Author},     
	pdfsubject={Subject},   
	pdfcreator={Creator},   
	pdfproducer={Producer}, 
	pdfkeywords={keyword1, key2, key3}, 
	pdfnewwindow=true,      
	colorlinks=true,       
	linkcolor=blue,          
	citecolor=blue,        
	filecolor=blue,      
	urlcolor=blue           
}

\usepackage{color}

\input{epsf}

\def\KAM{0.0154}

\def\Aubry{0.1386}

\begin{document}
	
	\title{Properties of phonon modes of ion trap
	       quantum computer in the Aubry phase}
	
	\author{Justin Loye}
	\affiliation{Institut de Recherche en Informatique de Toulouse, 
		Universit\'e de Toulouse, UPS, Toulouse, France}
	\affiliation{\mbox{Laboratoire de Physique Th\'eorique, IRSAMC, 
			Universit\'e de Toulouse, CNRS, UPS, 31062 Toulouse, France}}
	\author{Jos\'e Lages}
	\affiliation{Institut UTINAM, OSU THETA, CNRS, 
		Universit\'e Bourgogne Franche-Comt\'e, Besan\c con, France }
	\author{Dima L. Shepelyansky}
	\affiliation{\mbox{Laboratoire de Physique Th\'eorique, IRSAMC, 
			Universit\'e de Toulouse, CNRS, UPS, 31062 Toulouse, France}}
	
	\date{February 10, 2020}
	
	\begin{abstract}
		We study analytically and numerically the properties of phonon modes
		in an ion quantum computer. The ion chain is placed in a harmonic trap with an additional 
		periodic potential which dimensionless amplitude $K$ 
		determines three main phases
		available for quantum computations: 
		at zero $K$ we have the case of Cirac-Zoller quantum computer,
		below a certain critical amplitude  $K<K_c$ the ions are in 
		the Kolmogorov-Arnold-Moser (KAM) phase,
		with delocalized phonon modes and free chain sliding, and  above the critical amplitude
		$K>K_c$ ions are in the pinned Aubry phase with a finite frequency gap
		protecting quantum gates from temperature and other external fluctuations.
		For the Aubry phase, in contrast to the Cirac-Zoller and KAM phases, the phonon gap
		remains independent of the number of ions placed in the trap
		keeping a fixed ion density around the  trap center.
		We show that in the Aubry phase the phonon modes are much better localized
		comparing to the Cirac-Zoller and KAM cases.
		Thus in the Aubry phase
		the recoil pulses lead to local oscillations of ions while in other two
		phases they spread rapidly over the whole ion chains
		making them rather sensible to external fluctuations.
		We argue that the properties of localized  phonon modes and phonon gap
		in the Aubry phase provide advantages for the ion quantum computations
		in this phase with a large number of ions. 
	\end{abstract}

	%

	\maketitle
	
	\section{Introduction} 
	\label{sec1}
	The Cirac-Zoller proposal  1995 \cite{zoller} initiated active experimental investigations
	of a quantum computer
	with cold ions in a trap oscillator potential.
	In the same year, a  
	fundamental quantum logic gate had been experimentally realized
	\cite{monroe1995}, followed by
	realizations of entanglement, robust controlled-NOT quantum gates 
	and simple quantum algorithms
	with  two to four qubits 
	\cite{sackett2000,demarco2002,leibfried2003,schmidt2003}.
	This initial stage of development of ion quantum computers
	is reviewed in \cite{haffner2008,blatt2012}.
	The impressive progress in the quantum states control
	is highlighted in the Nobel lecture \cite{wineland2013}.
	
	In 2019, the ion quantum computers demonstrated 
	an impressive boost with realizations of several
	quantum algorithms with up to 11 qubits 
	\cite{kim10q,monroe2019nat,11q,monroefiedtheo,roos2019sci}.
	Hybrid classical-quantum algorithms have even been realized with 
	up to 20 qubits \cite{20qhybrid}. At present about 100 ions 
	can be trapped and kept routinely for hours \cite{100ions}.
	The review of present experimental status on ion quantum computing
	is given in \cite{ionrevmit}.
	
	This remarkable progress of ion quantum computing
	makes the scalablity problem of these computations of great actuality.  
	Indeed, the Cirac-Zoller 1995 proposal \cite{zoller} is not scalable
	for large $N$, since at fixed average distance between
	ions, the trap frequency $\omega_{tr}$ goes to zero 
	as $N \rightarrow \infty$, and thus the quantum gates
	become very slow. Also, the coupling between
	ion chain and the internal ion levels
	drops as $1/\sqrt{N}$. To avoid these scaling problems
	for large $N$, it was proposed to place ions in
	the Aubry pinned phase created by an additional periodic
	potential \cite{dlsqcion}. In the Aubry phase, the phonon
	excitations of the ion chain have an excitation gap
	independent of $N$ that is expected to protect
	the accuracy of quantum gates even for $N \rightarrow \infty$.
	
	The first theoretical study of ions in a 
	periodic potential had been reported in \cite{fki2007}.
	It was shown that this system can be locally described by 
	the Frenkel-Kontorova model 
	of particles connected by linear springs 
	and placed in a periodic potential of amplitude $K$ \cite{obraun}. 
	In this model, the equilibrium positions of the particles are described by 
	the Chirikov standard map \cite{chirikov}.
	The physics of the model  
	is understood from the properties of dynamical symplectic maps
	with invariant Kolmogorov-Arnold-Moser (KAM) curves at $K<K_c$
	and the fractal cantori replacing these KAM curves at $K>K_c$.
	The KAM phase corresponds to integrable dynamics
	while the cantori phase appears in the regime
	of chaotic dynamics \cite{chirikov,lichtenberg,meiss,aubry}.
	As proved by Aubry \cite{aubry}, the cantori state corresponds to
	the minimal energy of particles.
	The Chirikov standard map provides a local description of 
	various symplectic maps, and thus it describes a variety of 
	physical systems \cite{stmapscholar} 
	including  the properties of an ion chain
	in a periodic potential.
	
	At small potential amplitude $K$, the excitation of ions has 
	an almost acoustic spectrum starting from the trap frequency
	$\omega_{tr}$ which goes to zero as $N \rightarrow \infty$. However, when
	the optical lattice amplitude $K$  exceeds 
	a certain critical value $K_c$,
	the chain enters in the Aubry pinned phase
	with the appearance of an excitation gap $\omega_g$ 
	which is independent of the chain length and of the number of ions. 
	The main properties of this charge system are described in 
	\cite{dlsqcion,fki2007,ztzs,lagesepjd}. However, these studies
	mainly addressed the transport properties of charges, while for
	the analysis of the robustness of the quantum gates 
	it is necessary to understand the properties of 
	phonon excitations of the ion chain and its response to 
	the recoil momentum transfer
	from the laser pulses acting on internal ion 
	states. Here, we present the detailed study of
	such phonon properties performed with numerical simulations
	of the dynamics of 50 to 400 ions in the KAM and Aubry phases
	created by a periodic potential.
	
	We also note that the experimental investigations of
	cold ions in a periodic potential have been started in \cite{haffner2011,vuletic2015sci}
	with the first signatures of the Aubry transition reported with 5 ions
	by the Vuletic group \cite{vuletic2016natmat}.
	The experiments with a larger number of ions
	are now in progress \cite{ions2017natcom,drewsen}
	showing that the investigations of quantum computing in the Aubry phase
	are within the reach of state-of-the-art trapped cold ion techniques.
	
	The paper is organized as follows: in Section~\ref{sec2}, we give the system description,
	the properties of phonon modes are analyzed in Section~\ref{sec3},
	the properties of the recoil pulse propagation are described  in Section~\ref{sec4},
	the glassy properties of low energy ion chain configurations are
	depicted in Section~\ref{sec5}, discussion of results and conclusion are given in
	Section~\ref{sec6}. Appendix provides additional complementary material.
	
	\section{System description} 
	\label{sec2}
	
	The chain of ions 
	in a one-dimensional periodic potential is described by the following Hamiltonian
	\begin{equation}
	\begin{array}{cll}
	H &=& {\displaystyle\sum_{i=1}^N} \left( \displaystyle\frac{P_i^2}{2} + 
	\omega_{tr}^2 \displaystyle\frac{{x_i}^2}{2} - K  \cos x_i \right) + 
	\displaystyle\sum_{i > j} \displaystyle\frac{1}{|x_i - x_j|}
	\end{array}
	\label{eq:ham1d}
	\end{equation}
	where $x_i$ and $P_i$ are the conjugated coordinate and momentum of the
	$i$th ion, $\omega_{tr}$ is the frequency of the harmonic trap, 
	and $K$ is the amplitude of the optical array.
	The Hamiltonian
	is written in the following dimensionless units: the spatial period of the optical lattice is $\ell=2\pi$, the ion mass is $m=1$, the net atom charge is $|Q|=1$, and the Coulomb constant is $k_e=1$.
	In these atomic-type units, 
	the physical parameters are measured in
	units of length  $r_a= \ell/2\pi$,
	of energy $\epsilon_a = k_eQ^2/r_a=2\pi k_eQ^2/\ell$, of
	electric field strength
	$E_{a} = \epsilon_a/(|Q| r_a)$, 
	of particle velocity
	$v_a=\sqrt{\epsilon_a/m}$, and of time
	$t_a =  r_a \sqrt{m/\epsilon_a}$.
	For $\ell=1 \rm \mu m$ and $Q=e$, the dimensionless 
	temperature $T=0.01$ corresponds to the physical temperature
	$T = \pi k_eQ^2/(50\ell k_B) \approx 1$ Kelvin.
	The dimensionless Planck constant is
	$\hbar_{\rm eff}= (\hbar/|Q|)\sqrt{2\pi/(k_e\ell m)} \sim 10^{-4}$
	being very small for  $\ell \approx 1 \rm\mu m  $ and, e.g., ${^{40}}\rm Ca^{+}$ ions.
	Since $\hbar_{\rm eff}$ is very small 
	the ion dynamics can be considered as classical.
	
	The equilibrium static positions of ions in a periodic potential
	are determined by the conditions 
$\left\{\partial H/ \partial x_i =0, P_i=0\right\}$ for all $i=1,\dots,N$
	\cite{aubry,fki2007,lagesepjd}.
	In the nearest  neighbor approximation for
	interacting ions, these conditions lead to the symplectic map 
giving the recurrence relation of equilibrium ion
	positions $x_i$
	\begin{eqnarray}
	p_{i+1} = p_i + K g(x_i) \; , \; \; x_{i+1} = x_i+1/\sqrt{p_{i+1}} \; ,
	\label{eq:map}
	\end{eqnarray}
	where the effective momentum $p_i = 1/(x_{i}-x_{i-1})^2$
	is conjugated to $x_i$ and the kick function is
	$g(x) = - \sin x - (\omega^2/K) x$.
	In \cite{fki2007,lagesepjd}, it is shown that this map (\ref{eq:map}) describes well the actual equilibrium ion positions even when all
	interactions between ions are taken into account.
	Let us define the average ion density $\nu=N/L$ where $L$ is the number of spatial periods of the optical lattice.
	The map (\ref{eq:map}) can be locally linearized in order to obtain the Chirikov standard map. This procedure allows to find the Aubry transition threshold $K_c \approx 0.034 (\nu/\nu_g)^3$ as a function of the average ion density. Here, $\nu_g =1.618...$ is the golden mean density 
	at the central part of the chain \cite{fki2007,lagesepjd}.
	For $\nu = \nu_g$, the numerical data gives $K_c \approx 0.0462$
	being close to the theoretical value.

\section{Properties of phonon eigenmodes} 
\label{sec3}

The equilibrium ion positions are obtained numerically 
by the gradient method \cite{aubry}
and/or by the Metropolis algorithm described and already
used in \cite{fki2007,ztzs,lagesepjd}. 
The main minimization procedure
starts from the ground state configuration at $K=0$
with subsequent step by step increase of $K$ 
followed by energy minimization at each $K$ (see also discussion below). 
We consider the case
of ion density  $\nu \approx 1.618$ being fixed for the central part
of the chain near the trap minimum. This ion density is equal to the rotation number $\nu=N/L$, i.e., along the chain, $N$ ions are placed over $L$ optical lattice periods. The $\nu_g=\left(1+\sqrt{5}\right)/2=1.618...$ golden mean rotation number correspond to the most robust KAM curve (see e.g. \cite{lichtenberg,meiss}). This $\nu_g$ value
is taken as the average density of $1/3$-central part 
of the whole chain. It can be tuned by a change of $\omega_{tr}$.
Examples of ion positions
for KAM and Aubry phases are given 
in Appendix Figs.~\ref{figA1},~\ref{figA2} 
(see also \cite{fki2007,lagesepjd}). The phase space
of the map (\ref{eq:map}) and the equilibrium ion positions 
are shown in Appendix Fig.~\ref{figA3}.

\begin{figure}[t]
	\begin{center}
		\includegraphics[width=0.46\textwidth]{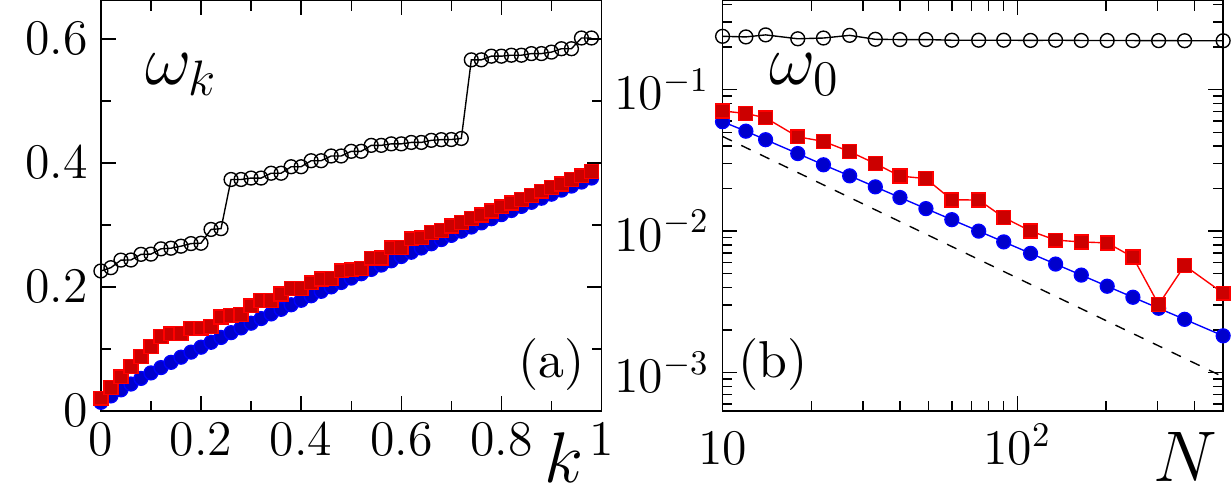}
	\end{center}
	\caption{\label{fig1}(a) Spectrum of phonon excitations $\omega_k$
		as a function of the scaled mode number $k=i/N$ ($i=0,\ldots,N-1$). 
		Here $N=50$, $\nu\approx\nu_g$, $\omega_{tr}=0.014$. 
		(b) Evolution of the lowest phonon mode $\omega_0$ as the number of 
		ions $N$ increases for $\nu\approx\nu_g$. 
		For both panels blue dots, red squares and black circles respectively correspond 
		to the Cirac-Zoller case ($K=0$), the KAM case ($K=\KAM\approx K_c/3$), and the 
		Aubry case ($K=\Aubry\approx3K_c$). The dashed line in panel (b) indicates the dependence
		$\omega_0 = 0.47/N^b$ with $b=1$, the fit of blue points for Cirac-Zoller case gives
		the exponent $b=0.892 \pm 0.00028$ (for the other cases the fit values of $b$ are:
		$b=0.743 \pm 0.023$ (KAM); $b=0.0132 \pm 0.0026$ (Aubry)).}
\end{figure}

\begin{figure}[t]
\begin{center}
\includegraphics[width=0.48\textwidth]{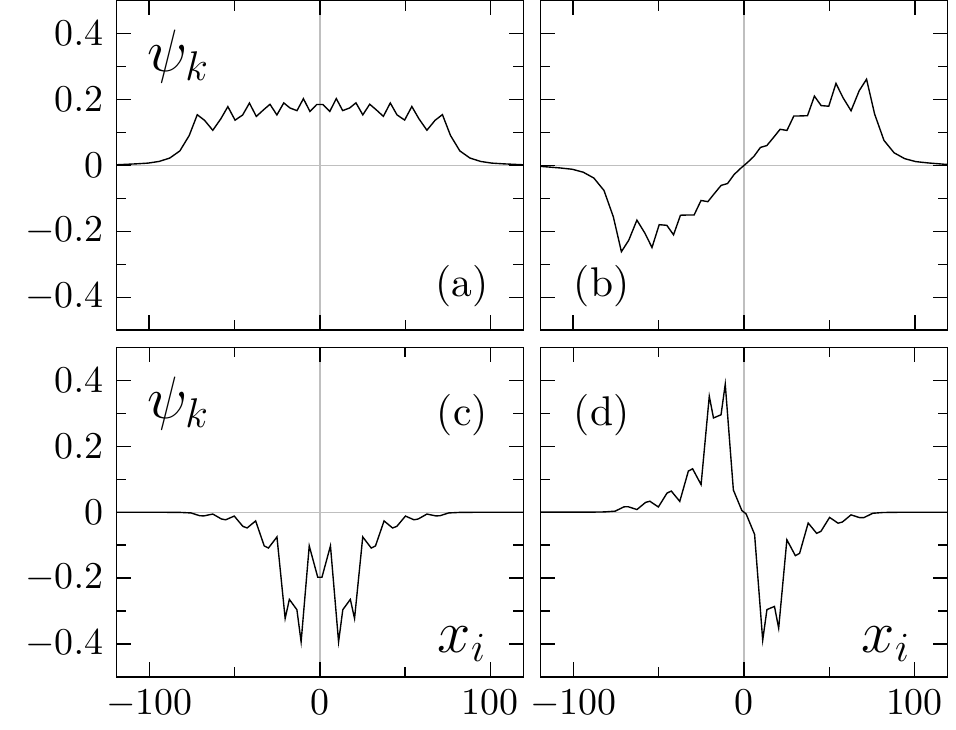}
\end{center}
\caption{\label{fig2}Amplitude of eigenmodes $\psi_k$ as a 
function of ion positions $x_i$ for the KAM phase at $K=\KAM\approx K_c/3$ 
(first row) and the Aubry phase at $K=\Aubry\approx 3K_c$ (second row), 
for $k=0$ (left column) and $k=1$ (right column). 
Here $N=50$, $\nu\approx\nu_g$, $\omega_{tr}=0.014$.}

\end{figure}

The Hamiltonian (\ref{eq:ham1d}) is linearized for small ion
oscillations near the equilibrium positions
and then the phonon spectrum and eigenmodes
are obtained by matrix diagonalization. Examples of phonon spectrum
$\omega(k)$ for the Cirac-Zoller case at $K=0$, and for the
KAM and the Aubry phases are shown in Fig.~\ref{fig1}. 
It is clear that for a large number of ions
the minimal phonon frequency $\omega_0$ goes to zero
for the Cirac-Zoller case and the KAM phase.
In contrast for the Aubry phase we have
an optical type phonon gap with $\omega_0 = \omega_g$
independent of the number of ions. Dependence of $\omega_0$ on $K$
is given in Appendix Fig.~\ref{figA4} 
(see similar cases in \cite{fki2007,lagesepjd}).

Examples of eigenmodes $\psi_k(x_i)$ with the two lowest frequencies 
$(k=0,1)$ are shown in Fig.~\ref{fig2}.
They clearly show that for the KAM case ($K<K_c$), and by extension for the Cirac-Zoller case ($K=0$), these modes
are delocalized over the whole chain
while for the Aubry phase ($K>K_c$) these modes are much more localized
 on a relatively small number of ions.
A more quantitative measure of the eigenmode spreading
 over the chain can be obtained from the inverse participation ratio (IPR)
defined for an eigenmode $\psi_k$ as
$\xi_k = (\sum_i | \psi_k(i) |^2)^2/\sum_i | \psi_k(i) |^4 $.
This quantity is broadly used in condensed matter for systems with 
disorder and Anderson localization (see e.g. \cite{akkermans}).

\begin{figure}[t]
	\begin{center}
		\includegraphics[width=0.48\textwidth]{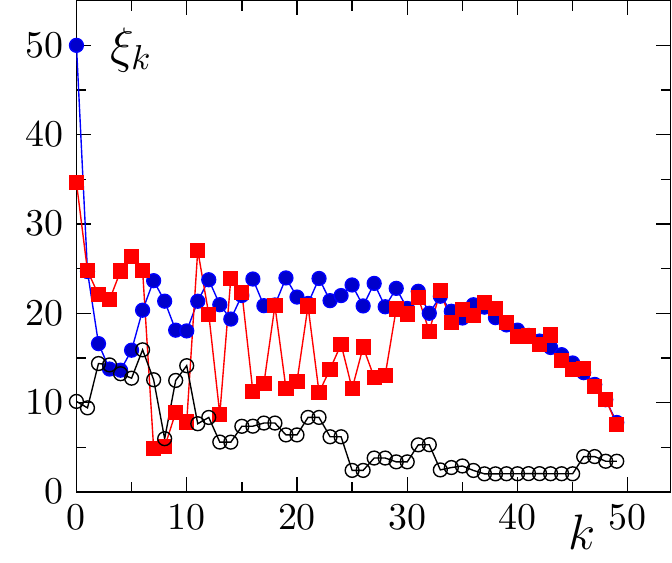}
	\end{center}
	\caption{\label{fig3}Inverse participation ratio $\xi_k$ as a function of mode index $k$ 
		for $K=0$ (blue dots), $K=\KAM\approx K_c/3$ (red squares), 
		$3K_c=\Aubry$ (black circles). Here $N=50$, $\nu\approx\nu_g$, 
		$\omega_{tr}=0.014$.}
\end{figure}

\begin{figure}[t]
\begin{center}
\includegraphics[width=0.48\textwidth]{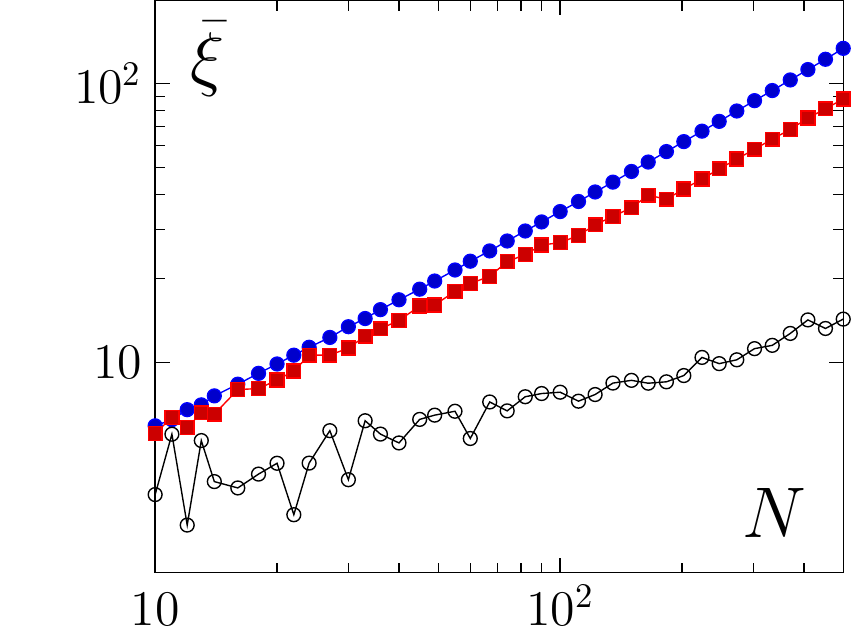}
\end{center}
\caption{\label{fig4}
Average inverse participation ratio ${\bar \xi}$ as a function of 
number of ions 
for the Cirac-Zoller case $K=0$ (blue dots), 
KAM case $K =0.0154 \approx K_c/3$ (red squares)
and the Aubry phase
$K=0.1386 \approx 3K_c$ (black circles). 
The algebraic fit $ {\bar \xi} = a N^b$ for
$\omega_{tr}=0.014$ gives respectively the value of exponent:
$b= 0.796 \pm 0.0029$ (Cirac-Zoller case),
$b= 0.696 \pm 0.0049$ (KAM case),
$b= 0.349 \pm 0.021$ (Aubry case).
}
\end{figure}

The IPR $\xi_k$ is shown in Fig.~\ref{fig3}
for the Cirac-Zoller case, the KAM phase and the Aubry phase.
The data shows that IPR values are significantly higher
for the Cirac-Zoller and KAM cases comparing to the Aubry case
with the average values $\bar{\xi} = \sum_k \xi_k/N = 20.001, 17.098, 6.148$
respectively for $N=50$. For higher $N$ values
we find respectively $\bar{\xi} =  48.411, 35.173, 8.248$ (for $N=150$)
and $\bar{\xi} = 86.282, 58.063, 11.685$ (for $N=300$).
Some higher eigenmodes are depicted in Appendix Fig.~\ref{figA5}.

The dependence of $\bar{\xi}$, averaged over all modes, 
on number of ions is shown
in Fig.~\ref{fig4}. 
This data clearly shows that in the limit of large number of ions
the Aubry case has much more localized modes
compared to the Cirac-Zoller and KAM cases.
Such localized modes are preferable for local quantum gates operation
since local modes interactions between modes and ions are stronger
and thus quantum gates are more rapid.
Indeed, the more the modes are localized, the more the ion displacement amplitudes are large, and consequently stronger is the coupling with the driving laser.
Also extended modes are expected to be more sensitive to various 
fluctuations.
Thus, we argue that the localization of phonon modes is expected 
to be more favorable 
for operation of quantum gates 
generated by laser pulses acting locally on 
internal states of a specific ion.

\section{Recoil pulse disintegration}
\label{sec4}

The study of the eigenmodes allows 
to understand the phonon properties of the ion chain.
However, the realization of quantum gates involves
the application of laser pulses which induce transitions
between one-qubit states (formed by internal ion levels)
and also induce interactions between qubits via ion recoil 
(see e.g. \cite{zoller,haffner2008,blatt2012}).

\begin{figure}[th!]
	\begin{center}
		\includegraphics[width=0.4\textwidth]{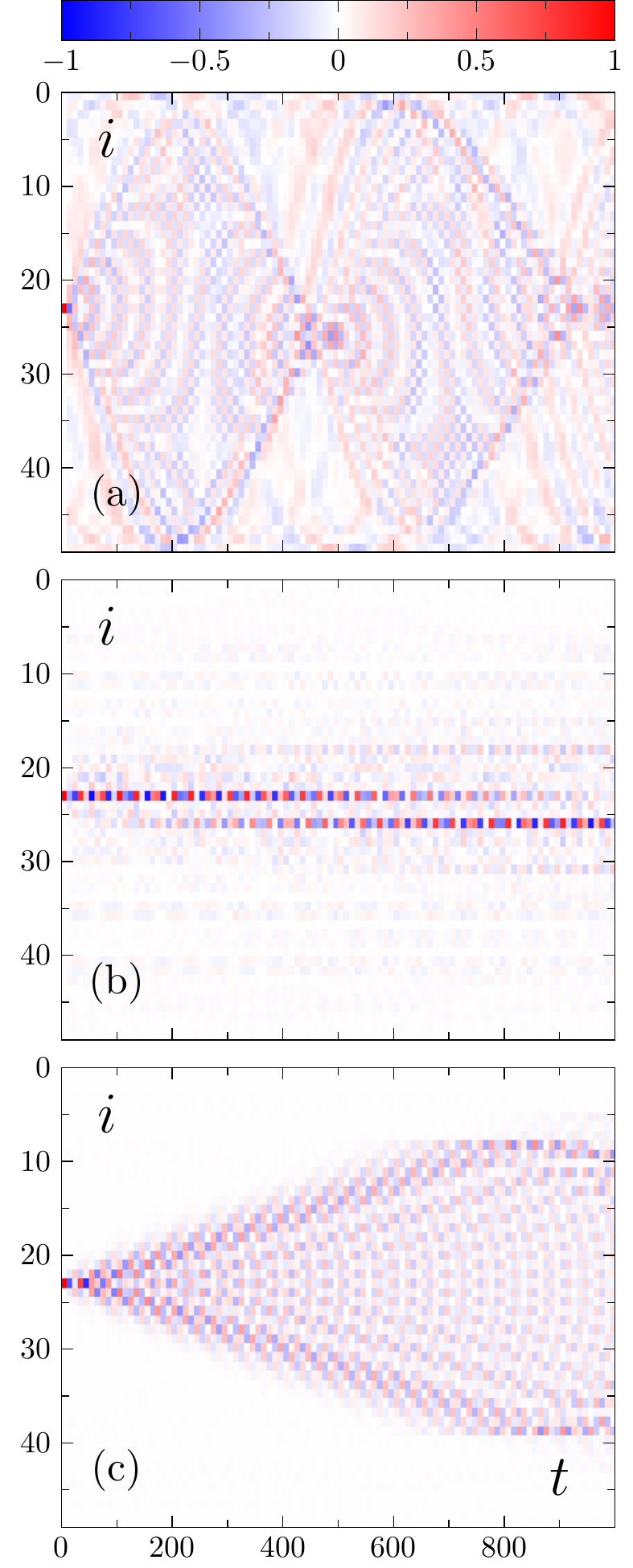}
	\end{center}
	\caption{\label{fig5}Recoil pulse spreading over ion chain in time. 
		The vertical axis shows ion number $i$, and the horizontal axis shows the dimensionless time $t$. 
		The initial momentum recoil $\delta P_0=10^{-3}$ is given to ion $i=23$, 
		for the Cirac-Zoller case $K=0$ (top panel), the
		Aubry phase at $K=\Aubry\approx 3K_c$ with $\nu\approx\nu_g$ (middle panel) and with  
		integer density $\nu\approx 1$ (bottom panel). Here $N=50$. The
		color gives the relative ion momentum $P_i/\delta P_0$.
	}
\end{figure}

In Fig.~\ref{fig5},  we present a numerical analysis of how an initial
small recoil impulse $\delta P_i$, given to a selected ion $i$ in the chain
at equilibrium (ions at rest), 
propagates through the chain.
The obtained results clearly show that
for the Cirac-Zoller case the initial impulse rapidly spreads through the whole ion chain
(top panel of Fig.~\ref{fig5}).
Indeed, in this case the recoil is transferred to the bus mode
leading to harmonic oscillations of ion chain.
This allows to obtain interactions between distant ion qubits. However,
the matrix element of such an interaction drops
as $1/\sqrt{N}$ (see Eq.(2) in \cite{zoller}).
Also the perturbative excitation of other modes 
leads to the propagation of perturbation all over the ion chain, 
as it is well visible in Fig.~\ref{fig5}. 
Such a perturbation disturbs the accuracy of quantum gates.
A similar situation is observed for the KAM phase
shown in Appendix Fig.~\ref{figA6}.  

The situation is qualitatively different for the Aubry phase.
Here the recoil impulse remains mainly localized 
between nearby ions with a rather weak
perturbation of other ions (middle panel of Fig.~\ref{fig5}).
Such a structure of recoil propagation
allows to create rapid two-qubit gates between nearby
ion qubits. Indeed, the time scale of gate operation $t_g$ 
is determined by the strong Coulomb interaction between nearby ions
and by the periodic potential amplitude. We can estimate 
the two-qubit gate operation time to $t_g \propto 1/\omega_g$. 
This time scale is independent of the number of ions $N$ in contrast to the 
Cirac-Zoller case. Of course, these Aubry two-qubit gates
in the Aubry phase  are local, acting between nearby ion qubits
while for the Cirac-Zoller case the bus mode
allows in principle to create a two-qubit gate
between distant ion qubits. However, the Cirac-Zoller
gates are slow in comparison with the Aubry gates. Moreover, 
the broad spreading of recoil pulse over the
whole chain  (Fig.~\ref{fig5} top panel)
will spoil rapidly the accuracy of such gates.

The advantage of the Cirac-Zoller proposal is the
ability to construct slow but long-range gates.
However, the coupling matrix element in such gates
drops as $1/\sqrt{N}$ and the minimal frequency of the gates also drops as 
$\omega_0 \propto 1/N$ (see Fig.~\ref{fig1}b).
Such gates are very slow and, thus, should be sensitive to
various fluctuations present in the system.
By contrast, in the Aubry phase, we can operate with gates 
acting mainly between nearby ions.
The corresponding strength of interactions between nearby
pinned ions is order of
unity in the chosen dimensionless units
(or $\epsilon_a$ since Coulomb interactions are
very strong between nearby ions).
These gates are rapid, local,
and thus they are expected to be much less sensitive to
fluctuations. Of course, there is a certain 
disadvantage in the local nature of these gates since 
one will need to perform about $N$ gates
to couple qubits on opposite ends of the chain,
but this feature is present almost for all
configurations of quantum computers 
where the gates are assumed to be mainly local.

Above, we discussed the Aubry phase with ion density $\nu \approx 1.618$.
At a first glance, one can suppose that it is possible 
to consider much simpler case at $\nu \approx 1$ when 
each lattice period contains only one ion
(if one neglects boundary effects of trap potential).
In fact, a configuration of a similar
type had been proposed by Cirac-Zoller in \cite{zoller2000nat}
where a periodic potential was assumed to be created by
equidistant microtraps. However, 
it was argued \cite{dlsqcion} that for $\nu =1$,
even with a high periodic potential amplitude,
Bloch waves will ballistically propagates
through the whole periodic crystal-like structure
according to the Bloch theorem \cite{bloch,kittel}.
The presence of such a Bloch wave at $\nu \approx 1$ is clearly
seen in Fig.~\ref{fig5} (bottom panel).
The wave propagates through the system till times
$t \approx 500$ leading to excitation of about half of all ions.
Finally for $t > 500$ this propagation becomes bounded
due to the presence of global harmonic trap with frequency
$\omega_{tr}$ which breaks periodicity of ion
chain at $\nu \approx 1$ due to correction
of ion local potential created by the global trap. Summarizing, for the $\nu=1$ case, a very large optical array amplitude $K$, inducing a strong pinning, produces an impulse propagation sharing similarities with the one associated to the Cirac-Zoller proposal.

The above result clearly shows that it is much more efficient
to place ions in the Aubry phase at irrational density $\nu$
where the recoil pulse propagation remains localized
in the vicinity of the initially kicked ion. We suppose that the localization of phonon modes
appears as a result of incommensurate 
ion density $\nu$ that leads effectively to 
a certain type of Aubry-Andre incommensurate model \cite{aubryandre}
where a transition from metal phase (delocalized)
to insulator phase (localized) appears with 
a variation of hopping matrix element between lattice sites.
In fact, the metal-insulator transition in the Aubry-Andre model
has been observed in experiment with cold atoms 
in an optical lattice \cite{inguscio,ibloch}.
However, a verification of direct relation
between localization of phonon modes
in the Aubry phase and the  Aubry-Andre model
requires further more extended investigations.

\section{Spin glass properties of low energy configurations of Aubry phase} 
\label{sec5}
The Aubry theorem \cite{aubry} guarantees that the cantori state 
with the fixed rotation number corresponding to the ion density 
provides the minimal energy ground state. However, in \cite{fki2007} is was shown 
that in fact there are exponentially many ion chain equilibrium configurations with
energies exponentially close to
the Aubry ground state. The number of such configurations
is growing exponentially with the increase
of the number of ions. Such a situation is similar to the case of spin glass systems
\cite{parisi} (see also  discussion in \cite{dlsqcion}). However, in the case of ion chain 
in a periodic potential there is no any
 external randomness and in a sense we have here a dynamical spin glass system
where the spin glass properties appear due to a dynamical chaos of the
related symplectic map. The existence of many low energy stable ion configurations means
that it is rather difficult to reach the Aubry ground state in real experimental conditions.
Above, we presented the results obtained with the minimization procedure 
with a step by step increase of $K$ amplitude 
starting each energy minimization procedure with a previous step.
We call this procedure as the gradual step $K$ increase minimization.
Such a procedure can be realized experimentally by a slow increase of $K$.
Other procedures can use also heating and cooling of ion chain
that can give other low energy configurations.
Thus, a real experiment will most probably operate not with the absolute minimal
Aubry ground state, but with a certain configuration with a local energy minimum
being very close to the energy of the Aubry ground state.
Consequently, it is important to unsure that these other configurations will be also
characterized by a phonon gap. With the minimization procedure developed in \cite{fki2007},
we determined about $N_s=200$ low energy configurations for $N=50$ and $N=150$.
The energy histograms of these configurations are given in
Appendix Fig.~\ref{figA7}. For these $N_s$ configurations
we checked that all of them have approximately the same phonon gap,
with an averaged width of $\omega_0$, and with an average variation
being rather modest 
$\delta \omega/\omega_g = [N_s^{-1}\sum_{N_s} (\omega_0 - \omega_{av})^2/{\omega_g}^2]^{1/2} \approx 0.046$ for $N=50$
(see Appendix Fig.~\ref{figA8}). 
We conclude that even if experiment will not operate with ions in 
the lowest energy configuration, still, the experimentally obtained low energy configuration
will have approximately the same properties as those discussed above.
Since the effective Planck constant is very small, the quantum tunneling
between these configurations will take an enormously large time as discussed in
\cite{fki2007,dlsqcion}. On such a timescale, the ions can be considered as 
practically frozen (with only small oscillations induced by recoil effects).

\section{Discussion.} 
\label{sec6}

The obtained results demonstrate 
a number of  significant advantages of a quantum computer
with ion qubit chain placed  in the Aubry phase  induced by an optical array.
This phase is characterized by a firm gap for phonon excitations
being independent of the number of ions. The phonon modes
in this phase are much stronger localized
on nearby ions comparing to the delocalized modes 
of the Cirac-Zoller proposal 1995 \cite{zoller}.
As a result, a recoil laser pulse 
remains mainly localized between nearby ions
that should allow the realization of a rapid two-qubit gates 
between nearby ion qubits.
These features open promising possibilities for
a scalable ion quantum computer in the Aubry phase.

However, future detailed investigations of 
ions in the Aubry phase are highly desirable in order to 
directly numerically model the quantum gates
taking into account both quantum nature of ion
motion and its internal states.

The experimental investigations
of ion quantum computer are within 
the reach of many experimental setups. Indeed, 
in \cite{drewsen}, 8 ions have been placed
in an periodic potential of depth of
$T \approx 25 \rm mK$. This depth is close to the Aubry transition
at $\nu \approx 0.38$ with $K_c \approx 4 \times 10^{-4}$
and  with a corresponding physical potential amplitude $V_A \approx 40 \rm mK$
for $\ell =1 \rm \mu m$ \cite{lagesepjd}.
Thus, we expect that the realization of two-qubit gates 
with ions in the Aubry phase can be performed with
available experimental conditions.

Finally, we note that the Hamiltonian (\ref{eq:ham1d}) 
can be also realized with electrons on a surface of liquid helium \cite{konobook}.
The experimental progress in a possible realization of such
a system is reported in \cite{kono1d,konstantinov}.
The efforts are also in progress \cite{kawakami} to 
realize a quantum computer with electrons on liquid helium
as proposed in \cite{platzman}.

\begin{acknowledgments}
We thank O.V. Zhirov for useful discussions and advises 
for various energy minimization procedures.
This work was supported in part by the Pogramme Investissements
d'Avenir ANR-11-IDEX-0002-02, 
reference ANR-10-LABX-0037-NEXT (project THETRACOM).
This work was also supported in part by the Programme Investissements
d'Avenir ANR-15-IDEX-0003, ISITE-BFC (project GNETWORKS), 
and in part by the Bourgogne Franche-Comt\'e Region
2017-2020 APEX Project. Also this work was supported 
by APR 2019 call of University of Toulouse and by 
Region Occitanie (project GoIA).
\end{acknowledgments}

\appendix*
\section{}
\label{appenda}
\setcounter{figure}{0}
\renewcommand{\thefigure}{A\arabic{figure}}
\renewcommand{\theHfigure}{\thefigure}

\begin{figure}[!t]
	\begin{center}
		\includegraphics[width=\columnwidth]{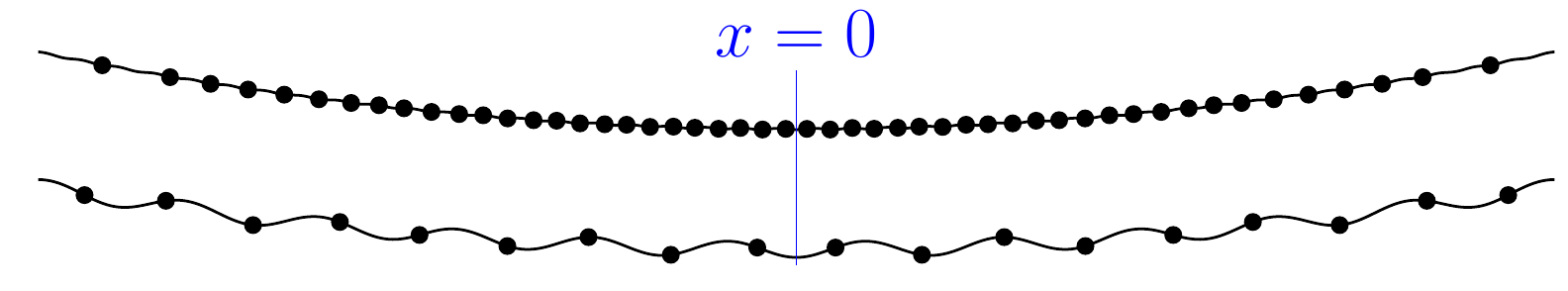}
	\end{center}
	\caption{\label{figA1}Chain representation in the KAM phase ($K=\KAM\approx K_c/3$). 
		The black line shows the potential $V(x)=\omega_{tr}^2 {x}^2/{2} - K  \cos x$ and the black dots 
		show the position of ions $x_i$. 
		On the first line the whole chain is represented while on the 
		second we zoom on the 17 central ions. 
		Here $N=50$, $\nu\approx\nu_g$, and $\omega_{tr}=0.014$. 
		We highlight the center of the harmonic trap with a blue vertical line.
	}
\end{figure}

\begin{figure}[!h]
	\begin{center}
		\includegraphics[width=\columnwidth]{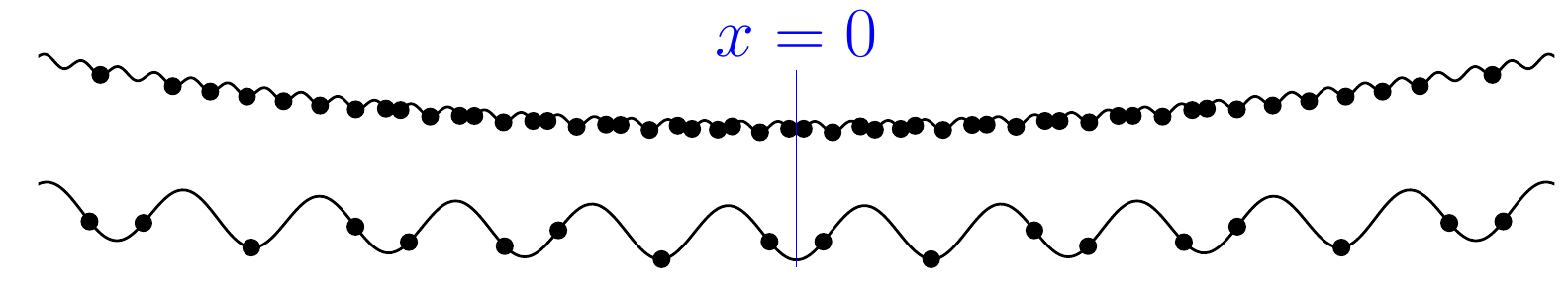}
	\end{center}
	\caption{\label{figA2}Same as Fig.~\ref{figA1} for the Aubry phase ($K=\Aubry\approx 3K_c$).}
\end{figure}

\begin{figure}[!h]
	\begin{center}
		\includegraphics[width=0.48\textwidth]{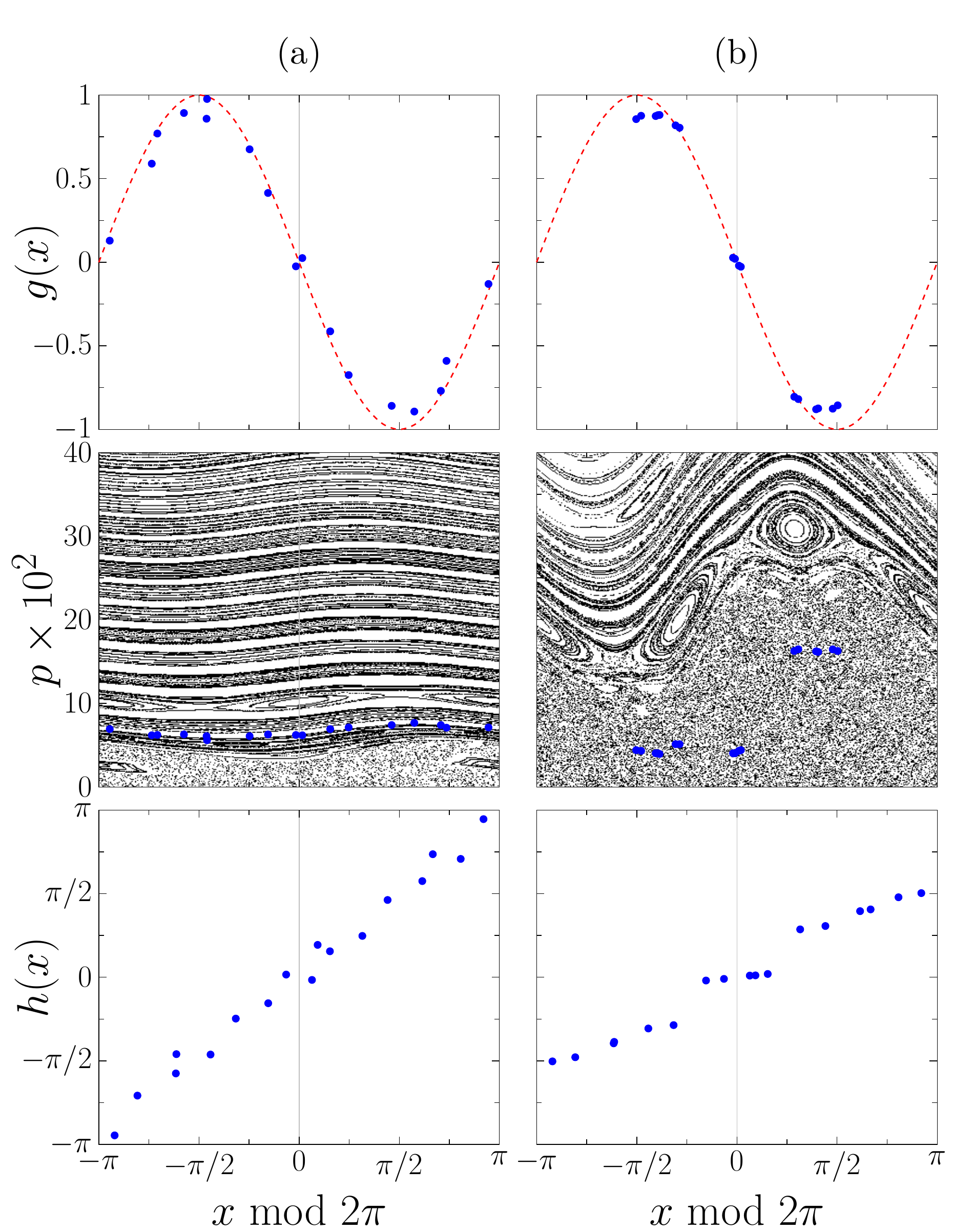}
	\end{center}
	\caption{Functions related to the ground state for $K=\KAM$ 
		(KAM phase at left column) and $K=\Aubry$ (Aubry phase at right column). 
		Top row: the kick function $g(x)=(p_{i+1}-p_i)/K$ obtained from the 
		ion  ground state positions
		$\{x_i\}_{i=N/3,..,2N/3}$ (blue dots); the dashed curve shows 
		the theoretical kick function $g(x)=-\sin x$. 
		Central row: phase space portrait of the 
		map \eqref{eq:map}; blue dots show the points obtained from the 
		equilibrium ion positions
		$\{p_i,x_i\}_{i=N/3,...,2N/3}$. Bottom row: the hull function $h(x)$ defined as 
		the positions of the ions $\{x_i\}_{i=N/3,...,2N/3}$ versus their positions 
		at $K=0$. Here $N=50$, $\nu=\nu_g \approx 1.618$, $\omega_{tr}=0.014$.}
	\label{figA3}
\end{figure}

\begin{figure}[!t]
	\begin{center}
		\includegraphics[width=0.32\textwidth]{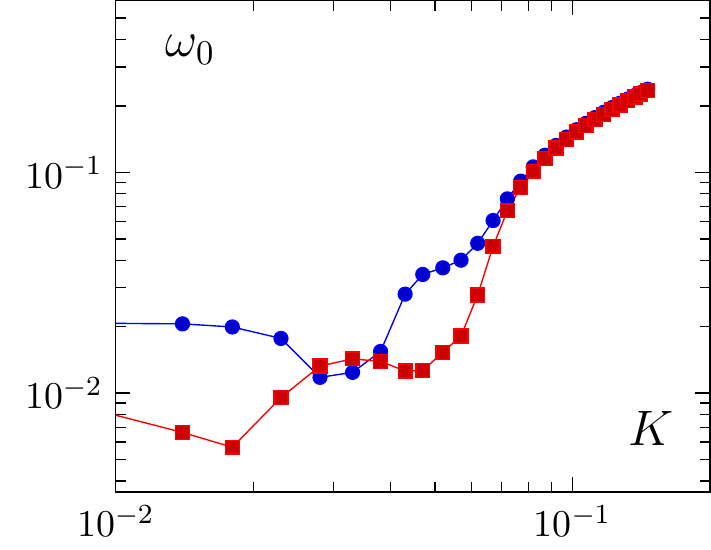}
	\end{center}
	\caption{Lowest phonon mode $\omega_0$ as the periodic potential 
		amplitude $K$ increases for $N=50$ (blue dots) and $N=150$ (red squares). Here $\nu=\nu_g$.}
	\label{figA4}
\end{figure}

\begin{figure}[!t]
	\begin{center}
		\includegraphics[width=0.48\textwidth]{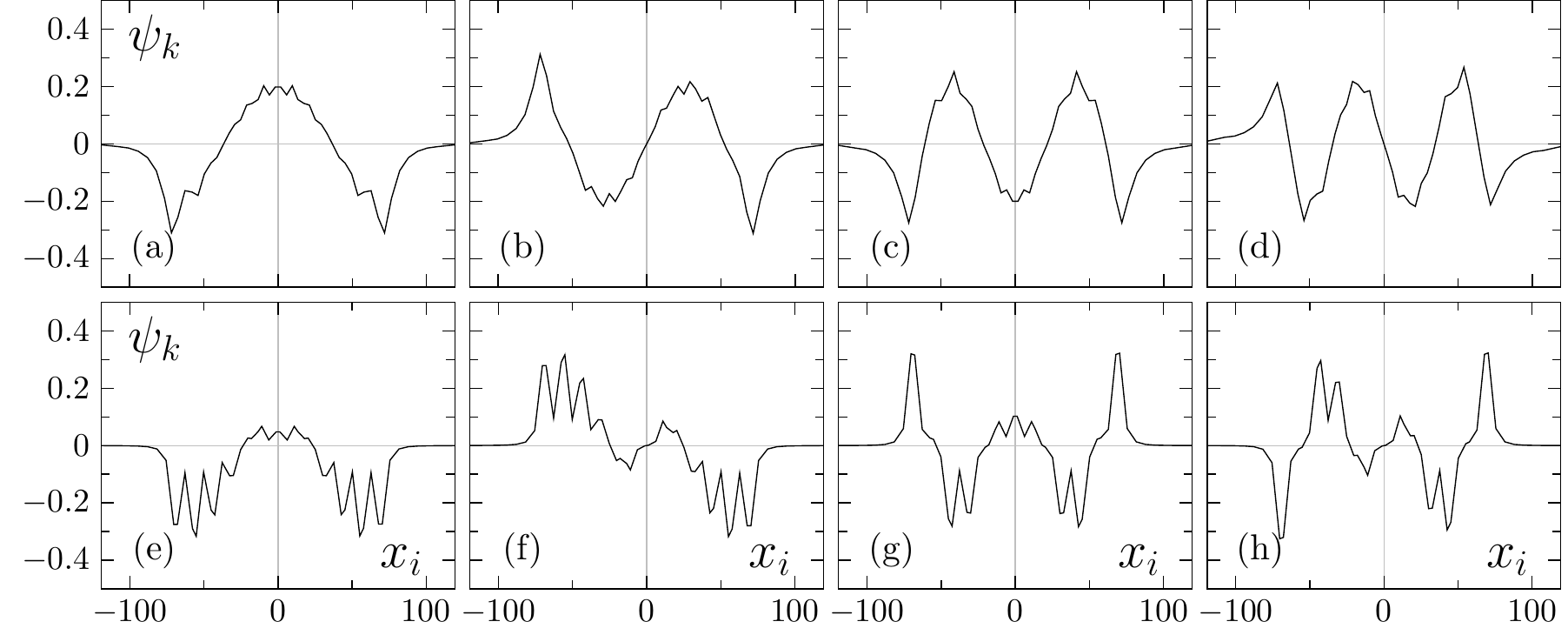}
	\end{center}
	\caption{Amplitude of eigenmodes $\psi_k(i)$ as a 
		function of ion positions $x_i$ for KAM phase at $K=\KAM\approx K_c/3$ 
		(first row) and Aubry phase at $K=\Aubry\approx 3K_c$ (second row) 
		for $k=2,3,4,5$ (columns left to right). 
		Here $N=50$, $\nu\approx\nu_g$, $\omega_{tr}=0.014$.}
	\label{figA5}
\end{figure}

\begin{figure}[t]
	\begin{center}
		\includegraphics[width=0.48\textwidth]{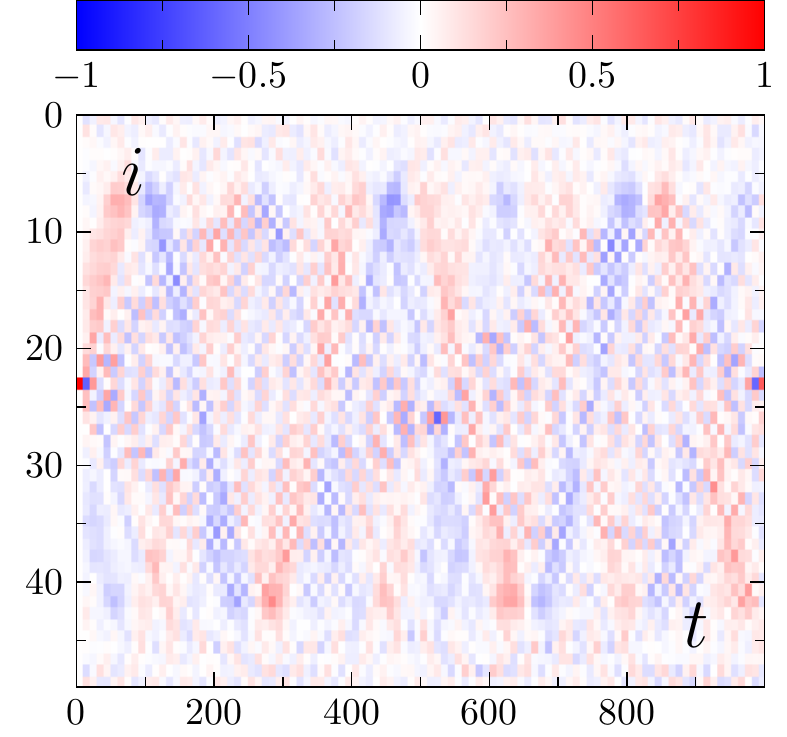}
	\end{center}
	\caption{Recoil pulse spreading over ion chain in time. The
		vertical axis shows ion number $i$, and the horizontal axis shows dimensionless the time $t$.
		The initial momentum recoil $\delta P_0=10^{-3}$ is given to ion $i=23$ in the KAM 
		phase $K=0.0154$. Here $N=50$, $\omega_{tr}=0.014$. The color gives the relative 
		ion momentum $P_i/\delta P_0$.}
	\label{figA6}
\end{figure}

\begin{figure}[!tb]
	\begin{center}
		\includegraphics[width=0.48\textwidth]{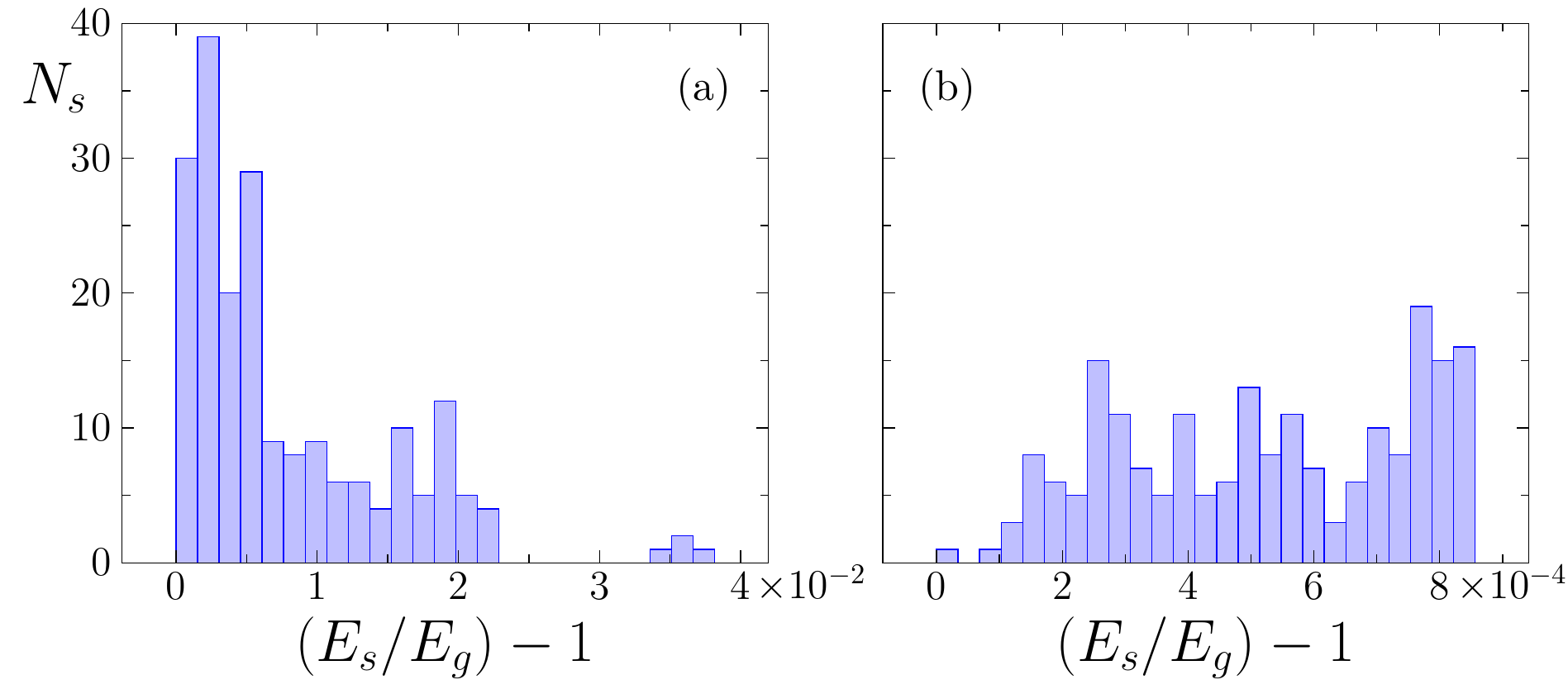}
	\end{center}
	\caption{\label{figA7}Distribution histogram 
		of fractions of  $N_s=200$ energy lowest stable ion configurations over
		their energy $E_s$ rescaled by the ground state energy $E_g$
		(shown in the vertical axis).
		Here $\omega_{tr}= 0.014$, $N=50$ (a) and $\omega_{tr}=0.005$, $N=150$ (b);
		$K=\Aubry$, $\nu \approx \nu_g$ in the central part 
		of the ion chain.
	}
\end{figure}

\begin{figure}[!h]
	\begin{center}
		\includegraphics[width=0.4\textwidth]{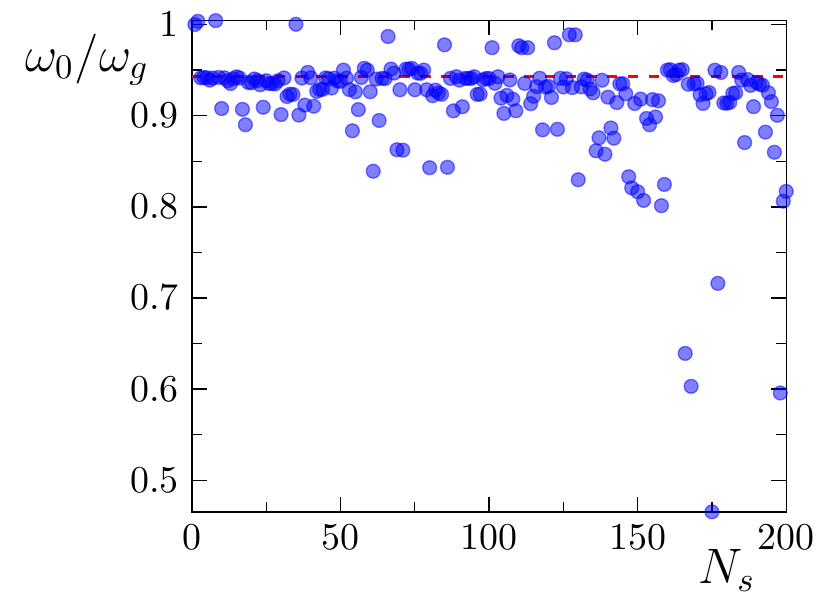}
	\end{center}
	\caption{\label{figA8}Distribution 
		of  $\omega_0/\omega_g$ 
		for the $N_s=200$ lowest energy configurations (ordered by energy
		starting from the ground state),
		where $\omega_0$ is the lowest mode phonon frequency 
		for the configurations and parameters of Fig.~\ref{figA7} (a),
		rescaling is done via the frequency $\omega_g=0.23969050023353308$ 
		of the ground state obtained via minimization procedure described in \cite{fki2007};
		dashed horizontal line marks the value of $\omega_0 = 0.2259555158207294$
		for the lowest energy configuration found by the gradual step $K$ increase 
		minimization used in the main part of the article.
	}
\end{figure}

Here we present supplementary  
figures Figs.~\ref{figA1}, \ref{figA2}, \ref{figA3}, \ref{figA4}, \ref{figA5},  \ref{figA6}, \ref{figA7}, and \ref{figA8}
complementing
the main text of the paper. We also add more detailed description of certain analytical 
results given in the main part of the article.


Figs.~\ref{figA1} and \ref{figA2} depict the equilibrium ion positions for KAM and Aubry phases.

The validity of the map description is illustrated in Fig.~\ref{figA3}.
Once the ground state of the chain obtained, the equilibrium positions of the ions, $\left\{x_1,\dots,x_N\right\}$, can be used with the map (\ref{eq:map}) to obtain the successive values $\left\{g\left(x_1\right),\dots,g\left(x_N\right)\right\}$ of the kick function $g(x)$ (blue dots in the top row panels of Fig.~\ref{figA3}). As expected these values lie on the top of the theoretical curve, $g(x)=-\sin x$, consequently validating the nearest neighbor approximation used to derive the map (\ref{eq:map}). This validation is obtained for both the KAM phase (left column) and the Aubry phase (right column). More precisely the kick function is $g(x)=-\sin x-(\omega^2/K) x$, but we can neglect the harmonic contribution since its value is at most of the order of $10^{-2}\pi$ for the KAM phase and of $10^{-3}\pi$ for the Aubry phase. The map (\ref{eq:map}) describes the one dimension dynamics of a fictitious particle which takes successive $\left(x,p\right)$ positions in the phase space. The middle row panels of Fig.~\ref{figA3} give the phase space portraits of the fictitious dynamics governed by the map (\ref{eq:map}). These phase space portraits are obtained by iterating the map (\ref{eq:map}) from many initial conditions $\left(x_0,p_0\right)\in\left[-\pi,\pi\right[\times\left[0,+\infty\right[$.
We observe that the ground state positions are located on an invariant KAM curve for $K<K_c$ (left column, KAM phase), and inside the chaotic component for $K>K_c$ (right column, Aubry phase).
The hull function $h(x)$ shown in bottom row panels of Fig.~\ref{figA3} is almost linear, $h(x)\approx x$, in the case of a sliding chain (left column, KAM phase), and $h(x)$ exhibits a devil's staircase typical for a pinned chain (right column, Aubry phase). More information can be found in \cite{ztzs,lagesepjd}.

The dependence of
the minimal phonon frequency
$\omega_0$ on potential amplitude $K$ is shown in Fig.~\ref{figA4}.
For $K<K_c$, we have $\omega_0$ being close to zero and
being practically independent of $K$.
In contrast for $K>K_c$, we have $\omega_0$ being independent of 
number of ions $N$
with $\omega_0$ growing with the increase of $K$.

A few high phonon  eigenmodes are shown in Fig.~\ref{figA5}.

The recoil pulse disintegration for the KAM phase is shown in Fig.~\ref{figA6}.

The energy distributions of low energy static ion configurations
for $N=50$ and $150$ are shown in Fig.~\ref{figA7}. They are obtained
with the minimization procedure described in \cite{fki2007}.
The energy of the Aubry ground state obtained with this procedure
is
$E_g=55.314$. It is slightly below the lowest energy
configuration obtained by the gradual $K$ step increase minimization
used for the figures of the main part of the article;
this procedure gives the energy
$E_s =  55.560$.
We note that both these values are significantly below the lowest configuration
energy obtained by the minimization procedure used in \cite{haffner2011}
which gives
$E_s=E_H=55.681$
(for $N=150$ we have respectively the energies
$216.956$,
$217.167$,
$218.045$).
We think that the minimization used in \cite{haffner2011}
effectively forbids ion hopping between potential minima
and due to that it is not able to find low energy configurations.
We think that it does not correspond to the reality of cold ion experiments
where ions can hop from one potential well to another.

In Fig.~\ref{figA8}, we show the variation of
the lowest phonon frequency $\omega_0$ of
the $N_s=200$ lowest energy configurations  for $N=150$
($\omega_0$ is rescaled by $\omega_g$
which is the frequency $\omega_0$ for of ground state
configuration).


\end{document}